\documentclass[
prd
,secnumarabic
,amssymb,nobibnotes, aps, prd]{revtex4}
\usepackage{amsmath}
\usepackage{graphicx}
\input{epsf}

\newcommand{\be}{\begin{equation}}    
\newcommand{\ee}{\end{equation}}
\newcommand{\beq}{\begin{eqnarray}}
\newcommand{\eeq}{\end{eqnarray}}
\newcommand{\beqn}{\begin{eqnarray*}}
\newcommand{\eeqn}{\end{eqnarray*}}

\def\nn{\nonumber}

\def\IL{\relax{\rm I\kern-.18em L}}

\begin{document}


\draft

\title{Stability of five-dimensional rotating black holes projected on
the brane}

\author
{E. Berti$^1$, K.D. Kokkotas$^1$ and E. Papantonopoulos$^2$}
\affiliation
{$^1$ Department of Physics, Aristotle University of Thessaloniki,
Thessaloniki 54124, Greece\\
$^2$ Department of Physics, National Technical University of Athens,
Zografou Campus, Athens 15780, Greece}

\date{\today}

\begin{abstract}

We study the stability of five-dimensional Myers-Perry black holes
with a single angular momentum under linear perturbations, and we
compute the quasinormal modes (QNM's) of the black hole metric
projected on the brane, using Leaver's continued fraction method. In
our numerical search we do not find unstable modes. The damping time
of modes having $l=m=2$ and $l=m=1$ tends to infinity as the black
hole spin tends to the extremal value, showing a behaviour reminiscent
of the one observed for ordinary 4--dimensional Kerr black holes.

\end{abstract}


\maketitle

\section{Introduction}

The existence of (compact or non-compact) extra dimensions in
brane-world scenarios offers a solution to the hierarchy problem,
exploiting the old idea that all known interactions except gravity are
confined on a three dimensional ``brane'' living in a
$(4+n)$--dimensional spacetime.  In some of the proposed scenarios,
the standard model matter and gauge degrees of freedom live on a three
brane within a flat space \cite{ADD}; alternative models exploit
properties of warped extra-dimensional geometries \cite{RS}.  A
consequence of the existence of large extra dimensions in brane-world
scenarios is that the true scale of quantum gravity may be
substantially lower than the four dimensional Planck scale $M_{Pl}$:
the fundamental Planck scale $M_{n}$ at which gravitational
interactions become strong can be of the order of a TeV. Then new and
interesting phenomena can be expected at energy scales comparable to
those currently available in accelerators \cite{LowEn}. Probably the
most striking and interesting possibility is the production of small
black holes by high-energy scattering processes, either in hadron
colliders such as LHC \cite{DL,GT} or in ultra-high energy cosmic rays
\cite{FSh}.

A black hole of mass $M$ produced at a lower Planck scale is quite
different from an ``ordinary'' black hole of the same mass
\cite{argyres}. Black holes relevant to experimental investigations in
hadron colliders or cosmic ray detectors have mass $M\ll M_{Pl}
\left(M_{Pl}/M_{n}\right)^{(n+2)/n}$ and horizon radius $r_h$ much
smaller than the scale of the extra dimensions $R$. Therefore they are
accurately described by the neutral, spinning high-dimensional Kerr
solutions of the $(4+n)$-dimensional Einstein action found by Myers
and Perry \cite{MP}.  When seen at distances $r\sim r_h$ the
$(4+n)$-dimensional spacetime is approximately flat, while at
distances much greater than the size of extra dimensions ($r\gg R$)
the usual, four-dimensional black hole solutions apply. The dynamics
of the brane is usually treated in the ``probe brane approximation'':
the only effect of the brane field is to bind the black hole, which
otherwise is treated as isolated, to the brane.

If small black holes are produced by collisions on the brane, they
will have non-vanishing angular momentum determined by the impact
parameter. In general, black holes in $D$ dimensions have $[(D-1)/2]$
rotation parameters. However, since the impact parameter is
non-vanishing only in directions along the brane, the angular momentum
is also lying within the brane directions. Hence we will only consider
multidimensional black hole solutions having a single spinning
direction. 

After production, black holes will first shed their hair in the
so-called ``balding phase'' emitting gauge radiation (standard model
fields on the brane and gravitons in the bulk), and then decay by
emission of Hawking radiation \cite{H}. The energy lost during the
black hole evaporation process is usually computed through a
semiclassical approach. However, it has recently been claimed that
taking into account the generalized uncertainty principle may have
dramatic effects on the energy loss \cite{CDM}. Quantum black holes
should be hotter, shorter-lived and have a smaller entropy than
semiclassical black holes of the same mass. Therefore less particles
would be emitted in the evaporation process with a larger average
energy, and the rate of black hole formation could be dramatically
suppressed.

It is generally believed that most of the emitted radiation will be
directed along the brane \cite{horowitz}. However, it has been pointed
out that this conclusion strongly depends on two approximations: the
assumption of Boltzmann statistics for the decay products, and the
assumption that the black hole mass does not change during the
radiation process. If these approximations are relaxed no
firm conclusion can presently be drawn on the energy loss into the
bulk, simply because the graviton greybody factors in higher
dimensions have not yet been computed \cite{C}.

The situation is even more complicated when one tries to include {\it
black hole rotation} in the picture. Simple estimates show that black
holes produced in high energy collisions are likely to have large
angular momentum \cite{GT,AFGS,IOP}. Black hole rotation may affect the
rates of black hole production in high-energy scattering processes in
two ways.

First of all, calculations in four dimensions suggest that the
probability of emission of gravitons by extremely rotating black holes
can be 100 times larger than the probability of emission of photons or
neutrinos \cite{page}. Therefore, even sticking to the ``standard''
semiclassical picture, bulk radiation may be comparable with the
radiation on the brane (or even be dominant) if the black hole rotates
fast enough. Emitting gravitons in the bulk the black hole can recoil,
move out of the brane and stop emitting brane-confined particles, so
that an observer on the brane will observe a virtual energy
nonconservation; this effect has been quantitatively studied in
\cite{frolov}.

A second motivation to include rotation in the picture is the
possibility of formation of ``black rings'', which have been suggested
to be thermodynamically favoured over multidimensional rotating black
holes at large rotation rates. Classical ``no-hair'' theorems show
that a stationary, asymptotically flat, vacuum black hole is
completely characterized by its mass and spin \cite{israel}. Similar
uniqueness theorems have recently been shown to hold for {\it static}
multidimensional black holes \cite{unique}. However, the situation for
{\it stationary} spacetimes is not so simple.  At present, only the
five--dimensional Myers--Perry metric with a single angular momentum
has been shown to be type--D \cite{desmet}. Furthermore, it has been
demonstrated by a counter--example that Myers--Perry black holes are
{\it not} unique among stationary higher dimensional spacetimes.

\begin{figure}[h]
\centering
{\includegraphics[angle=270,width=8cm,clip]{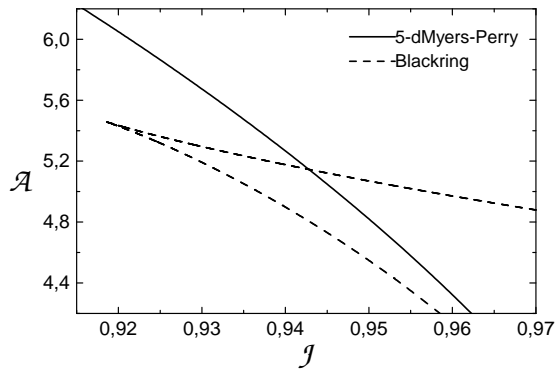}}
\caption{
Dimensionless horizon area (entropy) ${\cal A}$ of the 5-dimensional
Myers-Perry black hole (continuous line) and of the 5-dimensional
Emparan--Reall black ring solutions (dashed line) as a function of the
normalized angular momentum ${\cal J}$. The larger black ring solution
becomes entropically favoured when ${\cal J}={\cal J}_{ring}=0.9428$,
corresponding to a dimensionless horizon area ${\cal A}=5.146$ and to
an angular momentum parameter $a_*=2.828$.
}
\label{fig1}
\end{figure}

In fact, Emparan and Reall found a five-dimensional asymptotically
flat, stationary vacuum solution with an event horizon of topology
$S^{1}\times S^{2}$: a rotating ``black ring'', in which rotation
balances gravitational self-attraction \cite{emparan}.  Let us
introduce a dimensionless spin parameter, defined in terms of the
hole's angular momentum $J$ by
\be\label{Jdef}
{\cal J}=\sqrt{27\pi\over 32G}{J\over M^{3/2}}={a_*\over (1+a_*^2)^{1/2}},
\ee
where $a_*$ will be given as a function of the metric parameters in
the following.  Then the spin of a five-dimensional Myers-Perry black
hole is bounded from above \cite{note} by the condition ${\cal J}\leq
1$, and the extremal limit corresponds to $a_*\to \infty$.  On the
other hand the five-dimensional Emparan-Reall black rings, being
supported from collapse by rotation, have a spin which is bounded {\it
from below}, ${\cal J}>{\cal J}_{min}=\sqrt{27/32}\simeq 0.9186$. The
exact value of this lower bound has recently been determined by Elvang
\cite{E}. Elvang has found a charged extension of the Emparan-Reall
solution, showing that the charged ring's angular momentum can be made
arbitrarily small (charge supporting the ring against gravitational
self-attraction). However, we will limit the discussion to uncharged
rings, in which case the lower bound ${\cal J}_{min}>0$. For ${\cal
J}>{\cal J}_{min}$ {\it two} black ring solutions (one of which has
larger horizon area) coexist with the spinning Myers-Perry black hole.
As we show in figure \ref{fig1}, initially both black ring solutions
have a dimensionless horizon area (i.e., an entropy)
\be
{\cal A}=A/(GM)^{3/2}
\ee
which is smaller than the entropy of the Myers-Perry black hole having
the same rotation parameter. But when ${\cal J}>{\cal
J}_{ring}=0.9428$ the entropy of the larger ring becomes larger than
the entropy of the Myers-Perry black hole with the same ${\cal
J}$. Since the solution having larger entropy is the one expected to
be globally thermodynamically stable in the microcanonical ensemble,
Emparan and Reall conjectured that, as a five-dimensional black hole
is spun up, a phase transition from a black hole to a black ring
should occur.  However, such a phase transition involves a change in
the topology of the horizon \cite{Kol}, and it is not clear whether
this is possible classically \cite{maeda}.

Recently Ida {\it et al.} \cite{IOP} put forward arguments showing
that the maximum angular momentum of black holes produced in hadron
collisions is of the same order of magnitude of the minimum allowed
angular momentum for black rings, and argued that black ring
production should be even more favoured when the number of extra
dimensions is large. However, their suggestions are based on
order-of-magnitude estimates: the possibility of black ring formation
is not firmly grounded, and studies of the implications of such a
process are still in their infancy.

For the reasons listed above, a study of the stability of
multidimensional rotating black holes can be crucial for an
understanding of the possible production of ``black objects'' at LHC
or in cosmic ray observatories. This analysis involves the study of
black hole perturbations in more than four dimensions. 

Four dimensional black hole perturbations and the associated complex
oscillation frequencies, called quasinormal modes (QNM's), have been
extensively studied for more than thirty years, being of fundamental
importance in studies of gravitational radiation. For comprehensive
reviews see, eg, \cite{kokkotas}. However up to now there have been
just a few studies of multidimensional black hole perturbations. Only
recently (in the context of brane-world scenarios) have perturbations
of multidimensional black holes received more attention, and some
classes of multidimensional BH QNM's have been calculated
\cite{Konoplya}.

Besides the purely theoretical interest, studies of multidimensional
black hole perturbations are again motivated by phenomenological
reasons in scenarios involving black hole creation in accelerators or
cosmic rays. Indeed, according to simple dimensional arguments,
gravitational radiation should dominate over the emission of gauge
radiation in the balding phase \cite{GT}. The frequencies of this
gravitational radiation (or equivalently, the energies of the emitted
quanta) are nothing but the QNM frequencies. Some progress on the
computation of gravitational wave emission in multidimensional
spacetimes has been achieved by Cardoso {\it et al.} \cite{CDL}, who
extended some basic results in standard gravitational radiation theory
to higher dimensions.

Several studies of dynamical and thermodynamical stability of
higher-dimensional black holes have been carried out in the last few
years. For example, a quite detailed study of classical stability
properties of non rotating, higher dimensional black holes with
vanishing cosmological constant was carried out in \cite{GH}. Some of
these studies suggested that black branes are classically stable if
and only if they are thermodynamically stable \cite{stability}, but
recently some caution on this conjecture has been called for
\cite{Hartnoll}. A crucial step forward for studying classical
stability has been taken by Kodama and Ishibashi. Building on a
previously developed gauge-invariant perturbation formalism
\cite{KIS}, they were able to cast the perturbation equations of
higher-dimensional Schwarzschild black holes as second-order
Schr\"odinger--like perturbation equations \cite{KI}. There are three
such equations, corresponding to the different types of harmonics
(scalar, vector and tensor) used for the separation of the angular
dependence of the perturbations. In four dimensions, the radial
perturbation equation for scalar harmonics reduces to the Zerilli
equation, and that for vector harmonics reduces to the Regge-Wheeler
equation. Ishibashi and Kodama used this result to extend the
classical four-dimensional proofs of stability to higher dimensional
Schwarzschild black holes with or without cosmological constant
\cite{IK}. At present, only the stability of scalar perturbations of
non-asymptotically flat black holes is an open issue.

A similar, general treatment for {\it rotating} higher-dimensional
black holes is unfortunately still lacking.  Scalar perturbations of a
five dimensional Myers-Perry black hole were shown to be separable in
\cite{FS}; this result was later exploited in \cite{IUM} to compute
the scalar QNM's, and no instabilities were found.  The perturbation
equations for the Myers-Perry metric {\it projected on the brane} have
been derived, and again shown to be separable, in \cite{IOP}.  

In this work we investigate the stability of the induced metric on the
brane under scalar, electromagnetic and gravitational perturbations,
studying the generalized Teukolsky equation \cite{T} derived in
\cite{IOP}. We impose ingoing-wave boundary conditions at the black
hole horizon, outgoing-wave boundary conditions at infinity, and
compute the (complex) QNM frequencies in the five-dimensional case
using a continued fraction technique. We do not find any unstable
mode. Only as the black hole becomes extremal does the imaginary part
of gravitational modes with $l=m=2$ (and of electromagnetic modes with
$l=m=1$) tend to zero. Hence, no classical instability seems to set in
at the conjectured black hole/black ring phase transition. The
marginal instability in the extremal case is reminiscent of the one
observed in (four--dimensional) Kerr and
Reissner--Nordstr\"om--anti--de Sitter black holes.

The plan of the paper is as follows. In section \ref{contfrac} we show
how to extend Leaver's continued fraction method to the perturbations
of the on-brane projection of the Myers-Perry metric with a single
angular momentum. In section \ref{results} we present our main
numerical results. The conclusions and an outlook on possible
directions of research follow.

\section{Continued fraction method}\label{contfrac}

Our starting point is the induced metric on the three-brane in the
$(4+n)$-dimensional Myers-Perry metric \cite{IOP}. If only one of the
angular momenta is non-zero, the induced line element in
Boyer-Lindquist coordinates is given by
\be
ds^2=
\left(
1-{\mu \over \Sigma}r^{1-n}
\right)
dt^2+
{2a\mu \over \Sigma}r^{1-n}
\sin^2\theta 
dt d\phi
-
\sin^2\theta \left(
r^2+a^2+{\mu a^2 \sin^2\theta \over \Sigma}r^{1-n}
\right)
d\phi ^2-
{\Sigma\over \Delta}dr^2-\Sigma d\theta^2,
\ee
where $\Sigma\equiv r^2+a^2\cos^2\theta$ and $\Delta\equiv r^2+a^2-\mu
r^{1-n}$.  The parameters $\mu$ and $a$ are related to the mass $M$
and angular momentum $J$ by 
\be
M={(n+2)A_{n+2}\mu\over 16\pi G},\qquad
J={A_{n+2}\mu a \over 8\pi G},
\ee
where $A_{n+2}=2\pi^{(n+3)/2}/ \Gamma((n+3)/2)$ is the area of a unit
$(n+2)$-sphere.

The brane field perturbation equations for spin-$s$ fields separate
into an angular and a radial equation \cite{IOP}. The angular equation
is the same as in four dimensions, and the relevant boundary
conditions have been studied in \cite{L}. The radial equation can be
written in the following form:
\be\label{radD}
\Delta R_{lm,rr}+(s+1)[2r-(1-n)\mu r^{-n}]R_{lm,r}+V(r)R_{lm}=0,
\ee
where
\beq
V(r)&=&
\left[
(r^2+a^2)^2\omega^2-2am\omega \mu r^{1-n}+a^2m^2
+
is\{
am[2r+(n-1)\mu r^{-n}]
-\omega \mu [2r^{2-n}+(n-1)(r^2+a^2)r^{-n}]
\}
\right]\Delta^{-1}
\nn\\
&+&
2is\omega r-a^2\omega^2-sn(n-1)\mu r^{-n-1}-A_{lm}
\eeq
When $n=0$, equation (\ref{radD}) reduces to the radial equation
derived by Teukolsky for 4-dimensional Kerr black holes \cite{T}.  The
parameter $s=0,-1,-2$ for scalar, electromagnetic and gravitational
perturbations respectively, $a$ is the rotation parameter, and
$A_{lm}$ is an angular separation constant. In the non-rotating limit
the angular separation constant can be determined analytically, and is
given by the relation $A_{lm}=l(l+1)-s(s+1)$.

Quasinormal modes are characterized by waves which are purely ingoing
at the black hole horizon, and purely outgoing at infinity. Solving
the indicial equation, we find that ingoing wave solutions at the
horizon are such that $R_{lm}\sim (r-r_h)^{-s-i\sigma_+^{(n)}}$ as $r\to
r_h$, where
\be
\sigma_+^{(n)}=
{r_h\left[ \omega (r_h^2+a^2)-ma \right] \over
(n-1)(r_h^2+a^2)+2r_h^2}.
\ee
Consider now boundary conditions at infinity.  We want outgoing waves
there, so we assume $R_{lm}\sim r^x e^{i\omega r}$ as $r\to \infty$.
Then we get $x=-2s-1+i\omega$ when $n=0$, and $x=-2s-1$ for $n>0$. A
solution satisfying both boundary conditions can be expressed as
follows:
\beq\label{Rexp}
R_{lm}(r)=e^{i\omega r}
\left( {r-r_h\over r_h} \right)^{s-1+i\sigma_+^{(n)}}
\left( {r+r_h\over r_h} \right)^{-s-i\sigma_+^{(n)}}
\sum_k b_k
\left( {r-r_h\over r+r_h} \right)^k.
\eeq

From now on we will focus, for our numerical computations, on the
five-dimensional case ($n=1$). For a numerical treatment it is useful
to introduce dimensionless quantities $a_*=a/r_h$, $\omega_*=\omega
r_h$. The parameter $a_*$ is equivalent to the dimensionless angular
momentum ${\cal J}$, their relation being given by formula
(\ref{Jdef}). The $(4+n)$-dimensional horizon radius $r_h$ is related
to the mass and angular momentum parameters by
\be\label{horizon}
r_h=\left[{\mu\over(1+a_*^2)}\right]^{1/(n+1)}. 
\ee
This formula can be used to eliminate $\mu$, so that, measuring radii
in units of $r_h$, the radial equation only depends on the black hole
angular parameter $a_*$.

The coefficients $b_k$ in the series expansion (\ref{Rexp}) are
determined, when $n=1$, by a five-term recurrence relation (in
higher-dimensional spacetimes, the number of terms in the recurrence
relation grows) of the form:
\be
\alpha_k b_{k+1}+\beta_k b_k+\gamma_k b_{k-1}
+\delta_k b_{k-2}+\epsilon_k b_{k-3}=0,
~ (k=0,1,\dots)
\ee
where $b_{-k}=0$ and
$\{\alpha_k,~\beta_k,~\gamma_k,~\delta_k,~\epsilon_k\}$ are
second-order polynomials in $k$.
The coefficients of these polynomials are rather lenghty, but their
calculation is straightforward. Explicitly, they are given by
\beq
\alpha_k&=&
k^2+(-s-2i\sigma_+ +2)k
-s-2i\sigma_+ +1-\sigma_+^2+is\sigma_+ 
\\
&+&(isa_*m-is\omega_*-ma_*\omega_*+a_*^2\omega_*^2-is\omega_*a_*^2
-ma_*^3\omega_*)/2
+(\omega_*^2+\omega_*^2a_*^4+a_*^2m^2)/4
\nn\\
\beta_k&=&
-4k^2+(-2+6s+8i\sigma_+ +4i\omega_*)k
+2i\sigma_+ -6is\sigma_+ +\omega_*^2-\omega_*^2a_*^4
+is\omega_*a_*^2+is\omega_*
\nn\\
&-&a_*^2m^2-2s^2+2i\omega_*+2ma_*^3\omega_*
+4\sigma_+ \omega_*-a_*^2\omega_*^2-isa_*m+2ma_*\omega_*+4\sigma_+^2+s-1
-A_{lm}
\nn\\
\gamma_k&=&
6k^2+(-6-12i\sigma_+ -12s-8i\omega_*)k
+7s-8\sigma_+ \omega_*+8is\omega_*
+12is\sigma_+ -3ma_*\omega_*+4i\omega_*
\nn\\
&+&6i\sigma_+ +6s^2-3ma_*^3\omega_*
-6\sigma_+^2+a_*^2\omega_*^2+3+2A_{lm}
+(3\omega_*^2a_*^4+3a_*^2m^2-5\omega_*^2)/2
\nn\\
\delta_k&=&
-4k^2+(10s+8i\sigma_+ +10+4i\omega_*)k
-13s+isa_*m+4\sigma_+ \omega_*-9is\omega_*-10is\sigma_+ 
-\omega_*^2a_*^4-a_*^2\omega_*^2
\nn\\
&-&a_*^2m^2+2ma_*\omega_*
+\omega_*^2-6i\omega_*-10i\sigma_+ 
-6s^2+2ma_*^3\omega_*-is\omega_*a_*^2+4\sigma_+^2-7-A_{lm}
\nn\\
\epsilon_k&=&
k^2+(-3s-4-2i\sigma_+ )k
-\sigma_+^2+2s^2+6s+4+4i\sigma_+ +3is\sigma_+ 
\nn\\
&+&(is\omega_*-isa_*m+is\omega_*a_*^2-ma_*\omega_*-ma_*^3\omega_*
+a_*^2\omega_*^2)/2
+(a_*^2m^2+\omega_*^2+\omega_*^2a_*^4)/4
\nn
\eeq
The separation constant $A_{lm}$ can be obtained (once the other
parameters are fixed) solving an angular three-term continued fraction
relation identical to the one we get in four dimensions \cite{L}.

We can reduce the five-term radial recurrence relation to a three-term
recurrence relation using two Gaussian elimination steps: 
\beq
\alpha^{(1)}_0&=&\alpha_0,~\beta^{(1)}_0=\beta_0,
\\
\alpha^{(1)}_1&=&\alpha_1,~\beta^{(1)}_1=\beta_1,~\gamma^{(1)}_1=\gamma_1,
\nn\\
\alpha^{(1)}_2&=&\alpha_2,~\beta^{(1)}_2=\beta_2,~\gamma^{(1)}_2=\gamma_2,~\delta^{(1)}_2=\delta_2
\nn\\
\alpha^{(1)}_k&=&\alpha_k,~
\beta^{(1)}_k=\beta_k-\alpha^{(1)}_{k-1}\epsilon_k/\delta^{(1)}_{k-1},~
\gamma^{(1)}_k=\gamma_k-\beta^{(1)}_{k-1}\epsilon_k/\delta^{(1)}_{k-1},~
\delta^{(1)}_k=\delta_k-\gamma^{(1)}_{k-1}\epsilon_k/\delta^{(1)}_{k-1},
\qquad{k=3,4..}
\nn
\eeq
\beq
\alpha^{(2)}_0&=&\alpha_0,~\beta^{(2)}_0=\beta_0,
\\
\alpha^{(2)}_1&=&\alpha_1,~\beta^{(2)}_1=\beta_1,~\gamma^{(2)}_1=\gamma_1,
\nn\\
\alpha^{(2)}_k&=&\alpha^{(1)}_k,~
\beta^{(2)}_k=\beta^{(1)}_k-\alpha^{(2)}_{k-1}\delta^{(1)}_k/\gamma^{(2)}_{k-1},~
\gamma^{(2)}_k=\gamma^{(1)}_k-\beta^{(2)}_{k-1}\delta^{(1)}_k/\gamma^{(2)}_{k-1},
\qquad{k=2,3..}
\nn
\eeq
The treatment of higher-dimensional cases would involve recursion
relations having more terms (hence, a larger number of Gaussian
elimination steps) as the number of dimensions grows. There are no
conceptual complications in extending our method to higher dimensions,
but we intuitively expect the numerical stability of QNM searches to
decrease for increasing dimensionality.  The QNM boundary conditions
are satisfied when the following continued fraction condition on the
coefficients of the resulting three-term recurrence relation holds:
\be\label{CF}
0=\beta^{(2)}_0-
{\alpha^{(2)}_0\gamma^{(2)}_1\over \beta^{(2)}_1-}
{\alpha^{(2)}_1\gamma^{(2)}_2\over \beta^{(2)}_2-}\dots
\ee
Analytically equivalent conditions (which however are not, in general,
numerically equivalent to each other) are obtained inverting $j$ times
the previous continued fraction relation:
\beq\label{CFI}
&&\beta^{(2)}_j-
{\alpha^{(2)}_{j-1}\gamma^{(2)}_{j}\over \beta^{(2)}_{j-1}-}
{\alpha^{(2)}_{j-2}\gamma^{(2)}_{j-1}\over \beta^{(2)}_{j-2}-}\dots
{\alpha^{(2)}_{0}\gamma^{(2)}_{1}\over \beta^{(2)}_{0}}
={\alpha^{(2)}_j\gamma^{(2)}_{j+1}\over \beta^{(2)}_{j+1}-}
{\alpha^{(2)}_{j+1}\gamma^{(2)}_{j+2}\over \beta^{(2)}_{j+2}-}\dots
\qquad (j=1,2,\dots).
\eeq

In the four dimensional case ($n=0$) QNM frequencies scale with the
horizon radius, which by equation (\ref{horizon}) is proportional to
the mass, $r_h \sim \mu$. In five dimensions $r_h\sim \mu^{1/2}$; so,
following \cite{IUM}, we will give our results in terms of a
normalized frequency $w\equiv \mu^{1/2}\omega=(1+a_*)^{1/2}\omega_*$.

\section{Results}\label{results}

Using the techniques described in the previous paragraph, we have
carried out a numerical search for scalar, electromagnetic and
gravitational perturbation modes of the on-brane projection of
nonrotating ($a_*=0$) and rotating Myers-Perry black holes.

One of the most important results of our calculation is that {\it our
numerical searches showed no evidence for unstable modes} (notice that,
according to our conventions, the imaginary part of the modes is {\it
negative} for stable oscillations). Of course this is not a stability
proof, but such a result suggests that the on-brane projection of the
metric is stable with respect to linear perturbations, and maybe the
same applies to the ``full'' $(4+n)$-dimensional perturbations of
rotating Myers-Perry black holes.

In table 1 we give the fundamental QNM frequencies of the on-brane
metric for 5-dimensional, non rotating Schwarzschild black holes. For
each perturbing field we show the fundamental mode of the lowest
allowed multipole ($l=|s|$). We observe that the gravitational mode
with $l=2$ has the largest oscillation frequency and the smallest
imaginary part, followed by the electromagnetic mode with $l=1$ and by
the scalar mode with $l=0$.

\begin{figure}[h]
\centering
{\includegraphics[angle=270,width=8cm,clip]{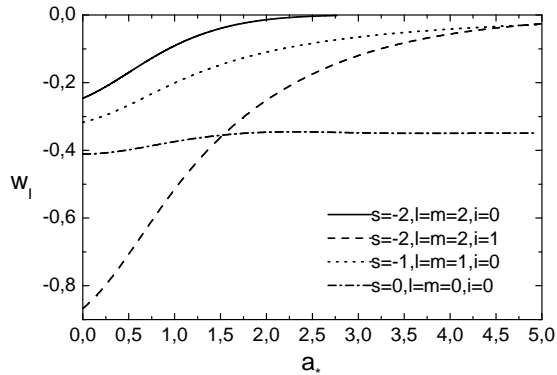}}
\caption{
Imaginary part of the normalized QNM frequency $w_I$ as a function of
$a_*$ for the fundamental gravitational mode with $l=m=2$ (continuous
line), the first gravitational overtone with $l=m=2$ (dashed line),
the fundamental electromagnetic mode with $l=m=1$ (dotted line) and
the fundamental scalar mode with $l=m=0$ (dot--dashed line). The index
$i$ labels the QNM overtone ($i=0$ for the fundamental mode, $i=1$ for
the first overtone). The imaginary part of the fundamental
gravitational mode with $l=m=2$ approaches zero when $a_*\gtrsim 3$,
and numerical calculations eventually break down. However, as
mentioned in the text, in a semilogarithmic plot the mode's imaginary
part shows an exponential behaviour ($w_I\sim e^{-c a_*}$) for small
$w_I$. The mode probably goes (marginally) unstable only when the
black hole becomes extremal ($a_*\to \infty$).
}
\label{fig2}
\end{figure}

As is well known from the study of perturbations of Kerr black holes,
rotation removes the degeneracy of perturbation modes having different
values of $m$ ($-l\leq m\leq l$). Following the trajectories of modes
having different values of $m$ in the complex plane, we observed that
modes having $l\neq m$ do not seem to go unstable for any rotation
rate. The results concerning modes with $l=m$ are more interesting.
In figure \ref{fig2} we plot the imaginary part of the normalized QNM
frequency $w_I$ as a function of the rotation parameter $a_*$ for the
fundamental scalar mode with $l=m=0$, for the fundamental
electromagnetic mode with $l=m=1$ and for the first two gravitational
modes with $l=m=2$. The fundamental scalar mode with $l=m=0$ does not
go unstable at all. However, one can see that the fundamental
electromagnetic mode with $l=m=1$ becomes marginally unstable ($w_I\to
0$) as $a_*\to \infty$. So do the gravitational fundamental mode
having $l=m=2$, and the first overtone. When the mode's imaginary part
becomes very small Leaver's continued fraction technique has
convergence problems (which are not a feature of the multidimensional
problem, and are well known to show up also in lower-dimensional
cases). Therefore we have difficulties in tracking the fundamental
gravitational mode with $l=m=2$ when $a_*>3$, and the imaginary part
becomes very small. This originally led us to suspect that this mode
goes unstable when $a_*\simeq 3$, suggesting a very interesting
connection with the possibility of a phase transition to the black
ring solutions found by Emparan and Reall. However, a closer
inspection shows that the imaginary part of the mode close to the axis
$w_I=0$ shows a behaviour of the type $w_I\sim e^{-c a_*}$, where $c$
is a constant. Using Leaver's method it is very hard to give a
definite answer to the question whether the $l=m=2$ gravitational mode
goes unstable at some finite value of $a_*$ or not. However, the
observed exponential behaviour suggests that the mode only becomes
marginally unstable as $a_*\to \infty$ (as do electromagnetic modes
and higher gravitational overtones). So (within the limits of our
on-brane calculation, which in some sense is intrinsically
four-dimensional) five-dimensional rotating black holes seem to behave
as they do in the four-dimensional case. Indeed, many years ago
Detweiler \cite{De} found a similar ``marginal instability'' for
ordinary Kerr black holes as they become extremal. However, Ferrari
and Mashoon \cite{FM} later showed that the mode amplitudes also go to
zero in the same limit, so the black hole does not undergo a real
instability.

\section{Conclusions and outlook}

We have computed QNM frequencies of the rotating Myers-Perry metric
projected on the brane using Leaver's technique. {\it We have not
found any unstable mode}, giving support to the stability of these
black holes, at least for perturbations ``confined on the
brane''. Furthermore, gravitational and electromagnetic perturbations
become unstable in the extremal limit (no such behaviour is observed
for the fundamental scalar mode having $l=m=0$).

Our result seems to suggest that {\it no classical instability}
(either in scalar, electromagnetic or gravitational modes) occurs at
the rotation rate corresponding to the conjectured black hole/black
ring phase transition.  
However, this is not in contrast with the Gubser--Mitra conjecture
\cite{stability}. Black holes are excluded from the scope of the
conjecture, which requires the existence of non--compact translational
symmetries. Furthermore, if an instability exists, it is likely to be
a Gregory--Laflamme type instability in the direction transverse to
the black hole rotation \cite{Reall}.

In any event, assessing the stability of the Emparan-Reall solutions
(and of eventual higher-dimensional generalizations) is an important
issue. The study of stability of higher-dimensional rotating black
holes and black rings ultimately calls for an extension of black hole
perturbation theory to higher dimensional spaces, in order to
determine which multidimensional ``black objects'' are classically
(and not only thermodynamically) stable.

Of course, since we did not consider perturbations of the {\it full}
five-dimensional metric, but only perturbations of its projection on
the brane, our analysis is (in some sense) intrinsically
four-dimensional. It is therefore necessary to study perturbations of
the full, $(4+n)$-dimensional metric. At present, to our knowledge,
there is no completely developed mathematical framework for such an
analysis for rotating black holes. We have also chosen to rely on the
so--called ``probe brane approximation''. The dimensional reduction
leading us to the on-brane metric presumably leads to some
modifications with respect to the full perturbation problem; in
particular, we ignored perturbations along the extra dimension, and we
cannot be sure that the massless gravitational mode is stably confined
on the brane in the strong field region of the solution. 

Moreover, the only known exact four--dimensional static spherically
symmetric black hole solutions on the brane give a gravitational
potential which goes as $r^{-2}$ for small $r$ and as $r^{-1}$ for
large $r$ in the case of the Randall--Sundrum model \cite{DMPR}. In
the case of a curved brane the gravitational potential goes as
$r^{-1}$ for small $r$ and as $r^2$ for large $r$ \cite{KPP}. In both
cases, such a behaviour cannot be recovered from a five--dimensional
Schwarzschild metric. The same sort of problem could arise in the
present case, changing the boundary conditions for QNM's and their
numerical value. The proof of such a statement is of course out of
reach at present, but we would expect the basic qualitative
conclusions of our paper (stability of the black holes, and marginal
instability in the limit of extremal rotation) to hold in the general
case as well.

Finally, we point out that electromagnetic and axial gravitational
QNM's of Reissner-Nordstr\"om anti-de Sitter black holes show a
similar marginal instability in the limit of extremal charge
\cite{BK}. In that case one has a topological phase transition
\cite{CEJM} and a notable change in thermodynamic properties
\cite{HHR}. There seems to be a close relationship between dynamical
properties of black holes (as described by their QNM's), their
thermodynamical description, and their topology. This relationship
definitely deserves to be explored in more depth.

\acknowledgments

It is a pleasure to thank M. Cavagli\`a, R. Emparan, J. P. Gregory and
H. S. Reall for useful comments and remarks. This work has been
supported by the EU Programme 'Improving the Human Research Potential
and the Socio-Economic Knowledge Base' (Research Training Network
Contract HPRN-CT-2000-00137).  E. P. wishes to thank the Physics
Department of the Aristotle University of Thessaloniki, where this
work was started, for hospitality.

\begin{table}
\centering
\caption{
Frequency of the fundamental modes for scalar ($s=0$), electromagnetic
($s=-1$) and gravitational ($s=-2$) perturbations of the non-rotating
black hole metric projected on the brane (in the non-rotating case, of
course, $w=\omega_*$). The first column gives the angular index $l$
(for each kind of perturbation we present results for the lowest
radiating multipole, so $l=|s|$). In the third and fourth columns we
give, respectively, the real and imaginary parts of the fundamental QNM
frequencies.
}
\vskip 12pt
\begin{tabular}{@{}ccc@{}}
\multicolumn{3}{c}{Fundamental QNM frequencies} \\
\hline
$l=|s|$  &$w_R$ &$w_I$ \\
\hline
0 & 0.27339 & -0.41091 \\
1 & 0.57667 & -0.31749 \\
2 & 0.80463 & -0.24622 \\
\end{tabular}
\label{table1}
\end{table}

\begin{references}
\bibitem{note}
In six or more dimensions, the black hole spin can be arbitrarily
large \cite{MP}.
\end{references}
\end{document}